\documentclass[12pt, a4paper]{article}
\usepackage{graphicx}
\newtheorem{theorem}{Theorem}

\newtheorem{definition}[theorem]{Definition}
\begin{document}

\title{Non-Cooperative Quantum Game Theory}
\author{Chiu Fan Lee\thanks{c.lee1@physics.ox.ac.uk}
\ and \ Neil F. Johnson\thanks{n.johnson@physics.ox.ac.uk}
\\
\\ Centre for Quantum Computation and Physics Department \\ Clarendon
Laboratory, Oxford University \\ Parks Road, Oxford OX1 3PU, U.K.}

\maketitle

\abstract{
The physical world obeys the rules of quantum, as opposed to classical,
physics. Since
the playing of any particular game requires physical resources, the question
arises as
to how Game Theory itself would change if it were extended into the quantum
domain. Here we
provide a general formalism for {\em quantum} games, and illustrate the
explicit application
of this new formalism to a quantized version of the well-known prisoner's
dilemma game.}
\\
\\
\noindent
{\bf Key words:}
quantum games, non-cooperative games, the prisoner's game

\newpage
\section{Introduction}
Quantum mechanics revolutionized physics a century ago.
During the past decade, quantum mechanics has extended its impact into the
fields
of information theory and computer science. By manipulating qubits---where
`qubit'
denotes the quantum equivalent of the single bit from classical information
theory---researchers have made some remarkable discoveries: perfectly
secure
cryptography is feasible \cite{BB84, Ek91, Be92},
certain computational tasks
can be performed more efficiently than their classical counterparts
\cite{Sh94, Gr96},
and teleportation and superdense coding have been proved possible
\cite{BB93, BW92}. The
key common element in each of these applications is information---and
information is
ultimately a physical quantity since it needs to be stored on, and
manipulated by, a
physical system. Games are no different in that their physical
implementation (i.e. the
playing of the game) will also require a physical system. In particular, the
actions of
players in games can ultimately be broken down into yes/no responses to
a series of specific questions
posed by an external referee. Since such a yes/no binary response has a
natural quantum
equivalent in terms of a qubit, game theory becomes an obvious candidate for
the
incorporation of quantum mechanical effects.

The study of quantum games started in 1999 \cite{Meyer}. Eisert, Wilkens and
Lewenstein
later proposed a quantized  version of the prisoner's dilemma game, claiming
that the
resulting `quantum game' resolves the prisoner's
dilemma \cite{EWL99}. This particular conclusion was later criticized
by Benjamin and Hayden
on the grounds that the quantum strategies considered were
limited in an unphysical way \cite{BH01}.
However the theoretical
framework introduced by Eisert and co-workers is unaffected
by this criticism, and moreover underlies most of the
subsequent research in quantum games \cite{Du02, BH01b,
Jo01, KJB01, LJ02n, LJ02, IT02, FA02}. In this paper, we
introduce a general formalism of quantum games based on Eisert {\it et al}'s
framework, 
and study 
the quantized prisoner's dilemma game as a specific example.

Before presenting a detailed discussion of quantum games, we start by
motivating the general formalism. In particular, we will
single out the essential elements in classical non-cooperative game
theory and then argue how these elements motivate the introduction of the
corresponding elements in the quantum version. We note that
from now on, all games considered will be non-cooperative and finite.
We will also restrict ourselves to two-player games, but will comment
on how this can be generalized.

Classically, any game is fully described by its
corresponding payoff matrix. For instance, the prisoner's dilemma game may
be represented by
the following payoff matrix which has two rows (labelled by 0 and 1,
for example) and two columns:
\begin{equation} 
\left[ \begin{array}{cc}
(3,3) &(0,5) \\(5,0) & (1,1)
\end{array} \right].
\end{equation}
We can then ignore the underlying motivation for the game in question, since
playing the game
becomes equivalent to picking a number corresponding to a particular row (or
column for the
second player) of the game matrix, e.g. 0 or 1.
Indeed, one of the successes of game theory may be seen as  the
incorporation of a utility
function into entries of a game  matrix, hence making a mathematical
treatment
possible. After the players have made their choices, e.g. by writing 0 or 1
on separate pieces
of paper, someone needs to collect together this information and distribute
the corresponding 
payoffs. Hence we assume the existence of a referee whose sole
purpose is to collate the choices made by the players and to assign
the corresponding payoffs. In short, playing a game constitutes an exchange
of information
between the players and the referee. The messages
exchanged can generally be thought to be encoded as bit-strings of fixed
length, hence
strengthening the information-theoretic theme. In the prisoner's dilemma
game, each player
only needs one bit to encode his/her choice: more generally, if there are
$n$ pure
strategies available for a specific player then $\log_2n$ bits are needed
to specify his/her
choice of strategy. Of course, there is nothing forbidding the
players from
picking their pure strategies randomly. This prompts
the mapping of each player's set of strategies to a multi-dimensional
simplex. Indeed, the
compactness and convexity of the strategy spaces and the multi-linearity of
the payoff
function are indispensable in proving the Nash Equilibrium Theorem and the
Minimax Theorem.

\section{Quantum games}
Given the many applications mentioned above in which qubits are manipulated
rather
than classical bits, an immediate approach to quantize classical games would
be to replace the
bits by qubits. A qubit may be regarded as a quantum system with two states
(a so-called
two-level system).  Physically, it may be represented
by the spin of an electron or the ground and excited states of an atom.
However this naive approach would not
give anything new. A more sophisticated approach was discussed
in Ref.~\cite{EWL99} and was later shown to be capable of
enriching the current
scope of classical game theory \cite{LJ02t}. Instead of players
manufacturing
their own qubits, the players in this approach operate on the qubits sent to
them by the
referee. Each player then sends his/her manipulated qubit back to
the referee. The players' strategy spaces are thus related to the spaces of
operators for the
qubits. This is similar to
the study of quantum error correction: there it is the vector space
of error operators on codewords which is of interest,
as opposed to the classical case in which it is the vector space of
codewords which
matters \cite{NC00}. The full procedure within this game-playing
scheme is therefore the following:
the referee first sends out sets of qubits to the players; the players
then operate separately on the qubits received; they then send the
resulting qubits back to the referee, who makes a
measurement on these qubits in order to determine the payoffs for each
player.
As we will show, quantum mechanics plays an essential role in this game by:

\begin{enumerate}
\item
restricting the feasible set of qubits
\item
restricting the physical operations available to the players, and
\item
restricting the physical measurements which can be performed by the referee.
\end{enumerate}

\noindent For classical games, we already have a clear picture of the
meaning of the three
restrictions above. Namely, (1) the messages exchanged are always encoded as
bit-strings;
(2) the operations allowed are restricted to tensor products of the bit-flip
($X$)
and the identity ($I$) operators, for example
$I\otimes X \otimes I (101) = 111$, and randomized ensembles of these
operators; (3) the
referee just `reads' the bit-strings received and assigns the payoffs
according to the payoff matrix of the game. In contrast, quantum
mechanics enforces far less
limitation on the vector space of qubits
and on the players' strategy spaces. On the other hand,
quantum mechanics does not allow perfect state estimation in general and
so the referee has limitations on how much he can learn from the
qubits. We will now give an axiomatic description on the elements
of quantum mechanics that concern us. A more complete treatment
may be found in Ref. \cite{NC00}. We note that all the matrices in this
paper can have complex numbers as entries.

\begin{definition}{(Description of qubits)}
A set of $n$ qubits
is described by a $2^n\times 2^n$ square matrix, $\rho$, such that
\begin{enumerate}
\item
{\rm tr}$(\rho)$=1,
\item
$\rho$ is a positive matrix.
\end{enumerate}
Any such $\rho$ is called a density matrix.
\end{definition}

\begin{definition}{(Description of physical operations)}
Given a density matrix $\rho$, any of the physically-implementable
operations on $\rho$
can be described by a set of square matrices, $\{ E_k\}$, such that
the $E_k$'s are of the same dimension as $\rho$ and $\sum_k
E_k^\dag E_k = I$. Moreover under any physical map $\{ E_k\}$, the
resulting density matrix will be $\sum_k E_k \rho E_k^\dag$
which is again a density matrix.

If $n$ qubits are divided into two subsets of $n_1$ and $n_2$ qubits,
and if each part is
operated upon separately, then any physical operation is
described by 
$\{ E_k\otimes F_l\}$, where the $E_k$'s are ($2^{n_1} \times
2^{n_1}$)-matrices
and the $F_l$'s are ($2^{n_2} \times 2^{n_2}$)-matrices
such that $\sum_k
E_k^\dag E_k = I$ and $\sum_l F_l^\dag F_l = I$. The
resulting density matrix will be $\sum_{k,l} (E_k
\otimes F_l) \rho (E_k^\dag \otimes F_l^\dag)$.
\end{definition}

\begin{definition}{(Description of physical measurements)}
A measurement with $L$ possible outcomes on
a density matrix $\rho$ corresponds to $L$
matrices, $\{ M_k\}$, of the same dimension as $\rho$,
such that $\sum_{k=1}^L M_k^\dag M_k =I$. The
probability of outcome $m$ is given by {\rm tr}$(M_m^\dag M_m \rho)$.
\end{definition}

Given the above definitions, we are now ready to introduce
a general theory of quantum games.
We will restrict ourselves first to two-player games.
We assume that the referee employs the measurement
$\{M_k \}$, assigning payoffs $a^{\rm I}_m$ and
$a^{\rm II}_m$ to players I and II
respectively if the outcome is $m$.
If
players I and II decide to use operations
$\{ E_k\}$ and $\{ F_k\}$ respectively, then the resulting state
$\pi$ will be given by $\sum_{k,l} (E_k
\otimes F_l) \rho (E_k^\dag \otimes F_l^\dag)$. Hence the payoff
for player I is
$\sum_{k=1}^L a^{\rm I}_k{\rm tr}(M_k^\dag M_k \pi)=
{\rm tr}[(\sum_{k=1}^L a^{\rm I}_kM_k^\dag M_k) \pi]$, with a similar
expression describing
the payoff for player II.
It is therefore convenient to denote $\sum_{k=1}^L a^{\rm I}_kM_k^\dag
M_k$ by
$R^{\rm I}$, and likewise for $R^{\rm II}$. We note that $R^{\rm I}, R^{\rm
II}$ and
the initial state $\rho$, define the game completely.

By treating the set of physical maps as a vector space, we may
fix a basis for it. We now suppose that $\{ \tilde{E}_\alpha \}$
form such a basis. Using
$\{ E_k = \sum_\alpha e_{k \alpha} \tilde{E}_\alpha \}$ and
$\{ F_l = \sum_\alpha f_{l \alpha} \tilde{E}_\alpha \}$, then the
payoff for player $j$ is
\begin{equation}
\label{1}
\sum_{k,l,\alpha, \beta, \gamma, \delta }
e_{k \alpha} \overline{e_{k \beta}} f_{l \gamma}
\overline{f_{l \delta}} A^j_{
\alpha \beta \gamma \delta}
\end{equation}
where $A^j_{\alpha \beta \gamma \delta} :=
{\rm tr} [R^j (\tilde{E}_\alpha \otimes \tilde{E}_\gamma)
\rho (\tilde{E}_\beta^\dag \otimes \tilde{E}_\delta^\dag)
]$. 
Letting $\chi_{\alpha \beta} = \sum_k e_{k \alpha} \overline{e_{k \beta}}$
and
$\xi_{\gamma \delta} = \sum_l f_{l \gamma} \overline{f_{l \delta}}$, then
$\chi$ and $\xi$ are positive Hermitian matrices
by construction. In quantum information, the above procedure is called
{\it $\chi$ matrix representation} \cite{NC00}.
We now rewrite Eq.~\ref{1} as follows:
\begin{equation}
\label{main}
\sum_{\alpha, \beta, \gamma, \delta }
\chi_{\alpha \beta} \xi_{\gamma \delta} A^j_{
\alpha \beta \gamma \delta}.
\end{equation}
We note that the tensors $A$s fully describes the game being played,
and the choices for $\chi$ and $\xi$ are limited by the laws of
physics. In particular, by denoting the set of allowable $\chi$ by
$\Omega$, the conditions described in definition~2
are transformed to the following conditions:
\begin{enumerate}
\item
$\Omega$ is a subset of the set of positive Hermitian matrices,
\item
For all $\chi \in \Omega$, then $\sum_{\alpha, \beta }
\overline{\chi_{\alpha \beta}} \tilde{E}_\alpha^\dag
\tilde{E}_\beta = I$.
\end{enumerate}
We can now see a striking similarity between static quantum games and
static classical finite games. The payoff for
a classical finite two-player game has the form
$\sum_{i,j} x_iA_{ij} y_j$ where $x,y$ belong to some multi-dimensional
simplexes and $A$ is a general matrix: the payoff for a static quantum game
is
$\sum_{\alpha, \beta, \gamma, \delta }
\chi_{\alpha \beta} \xi_{\gamma \delta} A_{
\alpha \beta \gamma \delta}$ where $\chi, \xi$ belong to some
multi-dimensional
compact and convex sets $\Omega$.
Indeed the multi-linear structure of the payoff function together with the
convexity and compactness of the strategy sets, are the essential features
underlying
both classical and quantum games. Indeed, we can
exploit these similarities in order to extend some
classical results into the quantum domain. Two immediate examples are the
Nash Equilibrium
Theorem and the Minmax Theorem \cite{LJ02t}.
We note that the classical strategy set, i.e.
a multi-dimensional simplex, and the quantum strategy set $\Omega$
cannot be made identical if the linearity of the payoff function is to be
preserved.
This is because there is no linear homeomorphism that maps $\Omega_k$ to a
simplex
of any dimension. In essence the positivity of $\Omega_k$,
i.e. the conditions $\chi_{\alpha
\alpha} \chi_{\beta \beta} \geq |\chi_{\alpha \beta}|^2$ for all
$\chi \in \Omega_k$, spoils this
possibility. 
Therefore, if we identify
$\Omega_k$ as some multi-dimensional simplex, we must lose linearity of the
payoff function. 
The structure of the strategy sets in the quantum case therefore
introduces new complexity to the study of finite games.

This entire analysis can easily be generalized to $N$-player games.
For instance, any particular $N$-player static game will have payoff
matrices of the form
$A^k = {\rm tr} [ R^k (\tilde{E} \otimes \cdots \otimes \tilde{E}) \rho
(\tilde{E} \otimes \cdots \otimes \tilde{E})]$ where we have omitted the
index summation for
clarity. 


\section{Quantized prisoner's dilemma game}
We now apply the formalism developed in the previous section, to discuss the
quantized prisoner's dilemma game introduced in Ref.~\cite{EWL99}.
Considering this game within our formalism, we are led to the following
forms for the initial
density matrix $\rho$ and the matrices $R^{\rm I}$, $R^{\rm II}$:
\begin{equation}
\rho= 
\left( \begin{array}{cccc}
1/2&0&0&-i/2 \\ 0&0&0&0 \\ 0&0&0&0 \\i/2&0&0&1/2
\end{array} \right),
\end{equation}
\begin{equation}R^{\rm I} =\left( \begin{array}{cccc}
2&0&0&-i \\ 0&5/2&5i/2&0 \\ 0&-5i/2&5/2&0 \\i&0&0&2
\end{array} \right)
\ \ \  \ 
R^{\rm II} =\left( \begin{array}{cccc}
2&0&0&-i \\ 0&5/2&-5i/2&0 \\ 0&5i/2&5/2&0 \\i&0&0&2
\end{array} \right).
\end{equation}
The motivation behind the above forms is as follows: if only
the bit-flip and the
identity operations are allowed, then the above game reduces to
the classical prisoner's dilemma game with the following game matrix
\cite{EWL99}:
\begin{equation} 
\left[ \begin{array}{cc}
(3,3) &(0,5) \\(5,0) & (1,1)
\end{array} \right]
\end{equation}
To perform concrete calculations, we identify the basis set
$\{ \tilde{E}_\alpha\}$ as
$\{ n_{[ij]} \}$ where $n_{[ij]} $ denotes an $n \times n$
square matrix such that the $(ij)$-entry is equal to 1 while
all other entries are equal to zero.
Denoting $\alpha$ by $ij$ and recalling the conditions on $\Omega$,
we have the following
restrictions on all $\chi \in \Omega: \sum_i \chi_{ijij}=1$,
$\sum_i \chi_{ijil}=0$ and
$\chi_{ijij}\chi_{klkl} \geq |\chi_{ijkl}|^2$,
where the first two sub-indices represent
$\alpha$ while the latter two represent $\beta$. A further calculation
shows that $n_{[ij]} \otimes n_{[kl]} = n^2_{[(i-1)n+k,(j-1)n+l]}$.
For arbitrary $R$ and $\rho$, we find the following:
\begin{equation}
\label{a}
A_{\underbrace{ab}_\alpha \underbrace{cd}_\beta
\underbrace{ij}_\gamma \underbrace{kl}_\delta}
=R_{((c-1)n+k),((a-1)n+i)} \times \rho_{((b-1)n+j),((d-1)n+l)}.
\end{equation}
We can therefore compute the $A$'s easily for the above quantized
prisoner's dilemma game. These are shown explicitly in Figure~1.

It can be seen that the matrices
contain imaginary numbers and negative numbers.
However there is no cause for concern: once we properly take the
conditions on
$\Omega$ into account, the resulting payoffs for the two players will
always lie between 0 and 5. From the form of the matrices,
one can also see that there is a Nash equilibrium
with payoff $2.5$ for each player: this corresponds to 
player I adopting the strategy $\chi^\ast$, where
$\chi^\ast_{0000}=\chi^\ast_{0101}=1$ and $\chi^\ast = 0$ for all other
entries and player II adopting the strategy $\xi^\ast$, where
$\xi^\ast_{1010}=\xi^\ast_{1111}=1$ and $\xi^\ast = 0$ for all other
entries.
One can check that the above $\chi$ and $\xi$
is contained in $\Omega$, and is hence
physically
implementable. We note that this is a Nash equilibrium
with the highest common payoff which is known in this game.

\begin{figure}
{\small
\caption{
The payoff matrices $A$ for the quantized prisoner's dilemma game.
The $i$-th row (column)
of $A$ corresponds to the first (last) four sub-indices
of $A$ representing the
binary expansion of $(i-1)$. For example, $A_{65}$ corresponds to
$A_{01010100}$ (c.f. Eq.~\ref{a}).}
\begin{center} Payoff matrix $A^{\rm I}$ \end{center}
\[
\left[
\begin{array}{rrrrrrrrrrrrrrrr}
1&0& 0& 0& 0& 0& 0& 0& 0& 0&\frac{5}{4}& 0& 0& 0& 0& 0\\
0& -i& 0& 0& 0&0& 0& 0& 0& 0& 0& -\frac{5i}{4}& 0& 0& 0& 0\\
0& 0& \frac{i}{2}& 0& 0& 0& 0& 0& -\frac{5i}{4}& 0& 0& 0& 0& 0& 0& 0\\
0& 0& 0& \frac{1}{2}& 0& 0& 0& 0& 0& -\frac{5}{4}& 0& 0& 0& 0& 0& 0\\
0& 0& 0& 0& i& 0& 0& 0& 0& 0& 0& 0& 0& 0& -\frac{5i}{4}& 0\\
0& 0& 0& 0&0& 1& 0& 0& 0& 0& 0& 0& 0& 0& 0& \frac{5}{4}\\
0& 0& 0& 0& 0& 0& -\frac{1}{2}& 0& 0& 0& 0& 0& \frac{5}{4}& 0& 0& 0\\
0& 0& 0& 0&0& 0& 0& \frac{i}{2}& 0& 0& 0& 0& 0& -\frac{5i}{4}& 0& 0 \\
0& 0& \frac{5i}{4}& 0& 0& 0& 0& 0& -\frac{i}{2}& 0& 0& 0& 0& 0& 0& 0\\
 0& 0& 0& \frac{5}{4}& 0& 0& 0& 0& 0& -\frac{1}{2}& 0& 0& 0& 0& 0& 0\\
\frac{5}{4}& 0& 0& 0& 0& 0& 0& 0& 0& 0& 1& 0& 0& 0& 0& 0\\
0& -\frac{5i}{4}& 0& 0& 0& 0& 0& 0&0& 0& 0& -1& 0& 0& 0& 0\\
 0& 0& 0& 0& 0& 0& -\frac{5}{4}& 0& 0& 0& 0& 0& \frac{1}{2}& 0& 0& 0\\
0& 0& 0& 0& 0& 0& 0& \frac{5i}{4}& 0& 0& 0& 0& 0& -\frac{i}{2}& 0& 0\\
0& 0& 0& 0&\frac{5i}{4}& 0& 0& 0& 0& 0& 0& 0& 0& 0& i& 0\\
 0& 0& 0& 0& 0& \frac{5}{4}& 0& 0& 0& 0& 0& 0& 0& 0& 0& 1
\end{array}
\right].
\]
\begin{center} Payoff matrix $A^{\rm II}$ \end{center}
\[
\left[
\begin{array}{rrrrrrrrrrrrrrrr}
1&0& 0& 0& 0& 0& 0& 0& 0& 0&\frac{5}{4}& 0& 0& 0& 0& 0\\
0& -i& 0& 0& 0&0& 0& 0& 0& 0& 0& -\frac{5i}{4}& 0& 0& 0& 0\\
0& 0& \frac{i}{2}& 0& 0& 0& 0& 0& \frac{5i}{4}& 0& 0& 0& 0& 0& 0& 0\\
0& 0& 0& \frac{1}{2}& 0& 0& 0& 0& 0& \frac{5}{4}& 0& 0& 0& 0& 0& 0\\
0& 0& 0& 0& i& 0& 0& 0& 0& 0& 0& 0& 0& 0& \frac{5i}{4}& 0\\
0& 0& 0& 0&0& 1& 0& 0& 0& 0& 0& 0& 0& 0& 0& \frac{5}{4}\\
0& 0& 0& 0& 0& 0& -\frac{1}{2}& 0& 0& 0& 0& 0& -\frac{5}{4}& 0& 0& 0\\
0& 0& 0& 0&0& 0& 0& \frac{i}{2}& 0& 0& 0& 0& 0& \frac{5i}{4}& 0& 0 \\
0& 0& -\frac{5i}{4}& 0& 0& 0& 0& 0& -\frac{i}{2}& 0& 0& 0& 0& 0& 0& 0\\
 0& 0& 0& -\frac{5}{4}& 0& 0& 0& 0& 0& -\frac{1}{2}& 0& 0& 0& 0& 0& 0\\
\frac{5}{4}& 0& 0& 0& 0& 0& 0& 0& 0& 0& 1& 0& 0& 0& 0& 0\\
0& -\frac{5i}{4}& 0& 0& 0& 0& 0& 0&0& 0& 0& -i& 0& 0& 0& 0\\
 0& 0& 0& 0& 0& 0& \frac{5}{4}& 0& 0& 0& 0& 0& \frac{1}{2}& 0& 0& 0\\
0& 0& 0& 0& 0& 0& 0& -\frac{5i}{4}& 0& 0& 0& 0& 0& -\frac{i}{2}& 0& 0\\
0& 0& 0& 0&\frac{5i}{4}& 0& 0& 0& 0& 0& 0& 0& 0& 0& i& 0\\
 0& 0& 0& 0& 0& \frac{5}{4}& 0& 0& 0& 0& 0& 0& 0& 0& 0& 1
\end{array}
\right].
\]}
\end{figure}

\section{Concluding remarks}
We have taken a brief tour through quantum games, and have shown that
quantum games are
indeed quite distinct from classical games. In particular, the strategy sets
are no longer 
simplexes and the payoff matrices admit complex entries.
These new features are certainly interesting from an academic point of view.
However the success of classical game theory lies in its applications. To
render
quantum game theory interesting, one must search for real-life scenarios
where
quantum games are useful. It turns out that such examples are not hard to
find: for example, 
multi-party communication schemes can naturally be envisaged as a
multi-player game.
Cryptography is another immediate example \cite{LJ02n}.  Moreover,
game-theoretic
language is well-suited to describe scenarios with multi-party interactions:
indeed
there are many examples of researchers discussing analogies between games
and quantum
systems long before the words `quantum games' were introduced \cite{MP95,
We98}. 
In addition, pursuing the underlying concept that information is physical
and that
physical systems can be seen as information-processors, one is led to the
idea that game
theory might even provide a novel interpretation of both classical and
quantum physics.
Such possibilities are likely to ignite future interest in game theory
within
the physical sciences both from the classical and quantum perspectives. In
short,
game theory is once again poised to extend its formidable range of
application -- however,
this time the application lies at the heart of fundamental science.

\end{document}